%% file: 0-main.tex
\documentclass[sigplan,twocolumn]{acmart}
\renewcommand\footnotetextcopyrightpermission[1]{}

\input{mac.tex}

\setcopyright{none}
\usepackage{epsfig,endnotes,url,color}
\usepackage{array}
\usepackage{multirow}
\usepackage{xspace}
\usepackage{verbatim}
\usepackage{wrapfig}
\usepackage[english]{babel}
\usepackage{amsmath}
\usepackage{caption}
\usepackage{balance}
\usepackage{tcolorbox}
\usepackage{booktabs}

\usepackage{graphicx}
\usepackage{subcaption}

\usepackage{hyperref}
\usepackage{hypcap}
\usepackage{xr}

\newcommand{\blackcircled}[1]{%
  \tikz[baseline=(char.base)]{
    \node[shape=circle,draw=black,fill=black,inner sep=0.2pt] (char) {\textcolor{white}{#1}};
  }%
}

\newcommand{\sys}{\textsc{EnclaveX}\xspace}

\begin{document}
	
\title{ \sys: End-to-End Confidential AI with CPU/GPU TEEs}

\author{Robert Schambach}\affiliation{TU Dresden}
\author{Quoc Do Le}\affiliation{STACKIT Cloud}
\author{Sergei Arnautov}\affiliation{Scontain GmbH}
\author{Christof Fetzer}\affiliation{TU Dresden}\affiliation{Scontain GmbH}

\settopmatter{printfolios=true}
\input{1-abstract}
\maketitle

\input{1-introduction}
\input{2-related-work}
\input{3-design}
\input{4-implementation}
\input{5-evaluation}
\input{6-conclusion}

\bibliographystyle{acm}
\bibliography{references}

\end{document}

\typeout{get arXiv to do 4 passes: Label(s) may have changed. Rerun}

%% file: mac.tex
\newcommand{\myparagraph}[1]{\smallskip \noindent{\bf {#1}.}}

\newcommand{\out}[1] {}

\newcounter{codeLineCntr}

\reversemarginpar
\newif\ifnotes
\notestrue

\newcommand{\punt}[1]{}

\renewcommand{\eqref}[1]{Equation~(\ref{eq:#1})}

\newcommand{\proc}[1]{\ifmmode\mbox{\textsc{#1}}\else\textsc{#1}\fi}

  \newcommand{\func}[1]{\ifmmode\mathrm{#1}\else\textrm{#1}fi} %

\newcounter{remark}[section]



%% file: 1-abstract.tex
\begin{abstract}
Large Language Models (LLMs) have rapidly proliferated, driving widespread adoption of AI applications. Most deployments rely on centralized infrastructures such as Microsoft Azure, Google Cloud, or AWS, requiring users to share sensitive data and training or fine‑tuning code. This dependence raises significant security and privacy concerns, as cloud providers must be trusted to ensure confidentiality and integrity.

Trusted Execution Environments (TEEs) e.g., Intel SGX/TDX, AMD SEV‑SNP, and ARM CCA have been introduced to mitigate these risks. More recently, NVIDIA has developed GPU TEEs (e.g., H100/H200), yet comprehensive evaluations of end‑to‑end workflows that integrate CPU and GPU TEEs remain limited. Critical aspects, including performance overhead, remote attestation, and security guarantees for AI/LLM applications, have not been sufficiently studied.

This paper addresses this gap by presenting an end‑to‑end workflow that combines CPU and GPU TEEs. We propose mechanisms to ensure confidentiality and integrity at both the VM level (via Intel TDX and AMD SEV‑SNP) and the application level, highlighting vulnerabilities such as Kubernetes administrators' ability to access confidential VM contents. Finally, we evaluate the performance overhead of our system using industry benchmarks, focusing on configurations that integrate Intel TDX with NVIDIA H200 GPUs.
\end{abstract}

%% file: 1-introduction.tex
\section{Introduction}

Cloud computing has revolutionized data processing~\cite{sgx_pyspark} including training/fine-tuning and inference AI/ML workload~\cite{securetf}, but it introduces profound security and compliance risks, particularly for sectors handling sensitive information like eHealth and finance. Traditional cloud setups expose data to threats from privileged insiders, malicious administrators, or supply-chain vulnerabilities, even during computation---``data in use'' remains a glaring blind spot in encryption paradigms that safeguard data at rest and in transit.

In eHealth, patient records and genomic data must comply with regulations like HIPAA, where breaches can erode trust and incur massive fines. Financial institutions grapple with PCI-DSS and similar standards, facing risks from model poisoning or data exfiltration during AI-driven fraud detection or algorithmic trading. Multi-tenant clouds amplify these issues, as shared infrastructure heightens the attack surface, while the rise of generative AI exacerbates concerns over intellectual property leakage and regulatory scrutiny under frameworks like the EU AI Act.

To this end, Trusted Execution Environments (TEEs) such as Intel SGX/TDX \cite{isgx, IntelTDXWhitepaper2022}, AMD SEV-SNP \cite{amd_sev_snp} or NVIDIA Hopper GPUs \cite{nvidia_tee} aim to reduce this attack surface by isolating application execution from the Cloud Service Provider (CSP) management software and infrastructure. %
Specifically, TEEs integrate well into the CSP orchestration and deployment software Kubernetes (K8s)~\cite{kubernetes}, allowing users to transparently deploy workloads inside of Confidential Virtual Machine (CVM) TEEs, e.g., using confidential containers~\cite{confidential_containers}. %
CVM-enclosed applications may further securely use an NVIDIA Hopper confidential GPU (cGPU) via a CVM-contained GPU driver, protecting CVM-cGPU IO and computation from the CSP.%

However, this state‑of‑the‑art approach still falls short of removing the entire CSP management stack from the application’s Trusted Computing Base (TCB), specifically, the K8s admin. In practice, a K8s admin retains full API privileges and can, for example, run {\em kubectl exec} to enter a confidential VM and access sensitive data, effectively operating as a root user inside the secure enclave.
To prevent this, secrets must be protected at the application layer. Keys should only be released to the application after successful attestation, similar to the Intel SGX model. With this design, even if an administrator manages to exec into a confidential VM, they still cannot access the application's encryption keys. %
We further disable memory dump features in the Guest kernel to prevent the K8s-admin to dump the application's memory within the CVM.%

In this paper, we introduce \sys, a comprehensive end‑to‑end Cloud Confidential Computing platform that reimagines secure AI/ML deployment. \sys integrates CPU TEEs such as Intel TDX for encrypted memory isolation, confidential GPUs such as NVIDIA H200 \cite{nvidia2024h200} for protected AI/ML acceleration, and SCONE~\cite{scone} runtime  for K8s‑native workflows. This combination enables remote attestation not only at the VM level through Intel TDX and NVIDIA H200, but also at the application layer by using SCONE.%

Together, these components enforce strong data sovereignty by verifying the integrity of code, data, and execution environments before releasing sensitive material. Organizations can therefore run high‑value workloads in public clouds while retaining fine‑grained control over access and compliance. The result is a scalable, high-performance framework that unifies HPC‑grade compute with robust security guarantees, enabling AI innovation without the constraints of traditional on‑premise silos.

%% file: 2-related-work.tex
\section{Related Work}%

\myparagraph{Process-based TEEs inside CVMs} %
Several works exist which create process-based TEEs inside AMD SEV-SNP CVMs \cite{wang_road_2025, zhao_vsgx_2022, ahmad_veil_2023}, protecting the process from the CVM OS using AMD Virtual Machine Privilege Level (VMPL). These systems also provide process‑level attestation in addition to CVM‑level attestation. %
In contrast, \sys is CVM‑agnostic and therefore compatible with Intel TDX, AMD SEV‑SNP, and Arm CCA. Whereas prior work relies on VMPL mechanisms, \sys uses a kernel module that is measured during CVM boot, avoiding dependence on any specific CVM instruction‑set extensions.%

Furthermore, while prior approaches restrict attestation to CPU‑based TEEs, \sys is, to our knowledge, the first to extend process‑based TEEs to confidential GPUs within CVMs. This brings SGX-style process isolation into CVMs with full confidential GPU support, enabling the process‑level TEE's security guarantees to extend across CPU-GPU boundaries and ensuring that I/O paths remain protected even from a root‑privileged K8s admin.

\myparagraph{Confidential GPU Benchmarks} %
A substantial body of work analyzes confidential GPU performance~\cite{chrapek_confidential_2025, ibarra_performance_2025, tan_performance_2024, yang_dissecting_2025, mohan_securing_2024, gu_nvidia_2025, zhuh200}, profiling the NVIDIA H100 and H200 confidential compute modes across diverse workloads. \sys builds on this foundation by presenting, to our knowledge, the first confidential Large Language Model~(LLM)‑inference benchmarks on an H200 cGPU that include full remote‑attestation measurements. Importantly, we provide empirical results quantifying the overhead of running the H200 in confidential compute mode inside a CVM compared to a native VM using the H200 in non-confidential mode. Furthermore, while prior studies focus on CVM-based TEE to cGPU execution paths, \sys is, to our knowledge, the first to evaluate process‑based TEE inside CVMs to cGPU workloads, extending the security boundary of process‑level TEEs inside CVMs into the cGPU domain.

%% file: 3-design.tex
\section{System Design Overview}
\subsection{Threat Model}
We design our system to withstand a highly capable adversary operating within complex cloud‑virtualized environments. In this model, the attacker controls the entire system software stack, including the operating system and hypervisor, and can carry out physical attacks such as memory probing. We assume that the K8s admin retains full API privileges and can invoke {\em kubectl exec} to access a confidential VM. We further assume an untrusted cloud network, enabling the adversary to drop, inject, replay, or modify packets and to manipulate routing. Collectively, these assumptions align with the classical Dolev–Yao adversary model~\cite{dolev-yao}.

Our threat model explicitly excludes side-channel attacks~\cite{sidechannel-attack1,sidechannel-attack2}, which are beyond the scope of this work. Nevertheless, the SCONE platform provides mitigation against L1-based side-channel threats~\cite{varys} and L2-based side-channel threats with AEX-Notify~\cite{aex-notify}. In addition, it is hardened against Iago attacks~\cite{checkoway2013iago}. To address Spectre-related vulnerabilities~\cite{Kocher2018spectre,chen2018sgxpectre}, we employ LLVM-based techniques such as speculative load hardening~\cite{Kocher2018spectre}.
Denial-of-service attacks are also out of scope, as they can be trivially executed by any infrastructure-controlling entity, such as the OS or hypervisor.

\subsection{Building Blocks}
\label{building_blocks}

\subsubsection{Trusted Execution Environments (TEEs)}
Serving as \sys{}'s security anchor, hardware-based TEEs including Intel SGX/TDX, AMD SEV-SNP, and ARM CCA, offer advanced ``in-use'' code and data integrity and confidentiality guarantees for cloud deployments. %
As TEEs extend hardware components, the extensions establish their root-of-trust in the vendor hardware instead of the Cloud-Service Provider (CSP) system software. %
Users may further assert the TEE hardware and enclosed-application authenticity and establish a secure communication channel thereto via remote attestation. %
Rather than isolating the execution completely from the CSP, TEEs offer strict management APIs to the system software. %
Hence, TEEs strike a balance of protecting their enclosed computation while natively fitting into the CSP deployment model~\cite{CCC2022Whitepaper}.%

Intel SGX introduced fine‑grained, process‑level isolation, but its adoption has been hindered by challenges such as complex development models, limited software compatibility, and performance overheads. More recent TEEs, such as Intel TDX, AMD SEV‑SNP, and ARM CCA, shift toward VM-level protection, securing entire Guest operating systems and unmodified applications.

The TEE model is increasingly extending into AI accelerators, including NVIDIA's H100/H200 and GB100/200 GPUs, as well as Graphcore IPUs. For example, NVIDIA's H100/H200 platforms support a unified TEE that spans both CPU and GPU, securing GPU memory and registers through PCIe‑level isolation and encrypting CPU–GPU communication channels to ensure confidentiality and integrity. Both the GPU and CPU TEEs protect their I/O over otherwise untrusted PCIe links using bounce buffers backed by hardware AES‑256‑GCM. To establish these protections, the TEEs negotiate a shared key via the Security Protocol and Data Model (SPDM) protocol~\cite{nertney2023confidential}.

\subsubsection{SCONE: Confidential Computing Platform}
{\bf Shielded Execution:}
At the software core, SCONE provides a shielded execution framework that enables unmodified applications to run inside TEE enclaves (e.g., Intel SGX/TDX). Built on Intel SGX/TDX and compatible with emerging confidential GPU technologies, SCONE encrypts data and code in use, shielding them from host OS, hypervisors, or cloud providers. 
Its hallmark is a minimal Trusted Computing Base (TCB) enabling granular trust at the microservice level. This isolates individual components, drastically reducing the attack surface compared to full-VM TEEs, while facilitating straightforward audits through remote attestation reports that verify enclave integrity without exposing contents.
SCONE's architecture supports unmodified applications, wrapping them in secure wrappers that handle encryption keys and attestation seamlessly.

\subsubsection{Remote Attestation}%
\sys{}'s ability to prove its trustworthiness via remote attestation is integral to the system's security. %
Following the Remote Attestation Procedures Architecture (RATS)~\cite{rfc9334} and Trusted Execution Environment Provisioning (TEEP) Architecture~\cite{rfc9397}, remote attestation enables an Attester to provide a set of verifiable claims about itself to a Relying Party, which may consider if the Attester is trustworthy. %
In the context of TEEs, these claims typically include TEE properties, the TEE's manufacturer, or which trusted application is running in the TEE. %
Finally, the relying party forwards the evidence to a Verifier, which validates the evidence with possibly externally provided endorsements, reference values, or an appraisal policy for the evidence. %
The Verifier then returns the attestation results to the relying party, which may again check the results with an additional appraisal policy. %
Notably, the exact order of the attestation data flow may vary.%

\sys{}'s attestation builds upon CVM, namely Intel TDX and AMD SEV-SNP, and cGPU attestation. %
To this end, we introduce necessary background for each of these distinct attestation procedures.%

\myparagraph{CVM} %
CVM attestation enables a relying party to ensure a CVM is running in an expected state on genuine CVM-enabled hardware. %
The CVM attestation flows follow a similar schema; the relying party receives a CPU-signed attestation report containing boot measurements. %
The relying party then verifies the report signature using vendor certificates and checks the report-contained measurements.%

\textit{Intel TDX.} %
For TDX attestation, the TD Attester provides evidence to the relying party in form of a signed attestation report, termed a quote. %
The TD thereby derives this quote by initially obtaining a report from a host-included TDX Module and forwarding this report to a signing host-located yet TD-external Quoting Enclave~(QE). %
This QE then signs the report with an Intel Provisioning Certification Key (PCK)-signed Attestation Key (AK). %
The QE returns the resulting quote to the TD, which again forwards this signed evidence to the relying party. %
The relying party may then act as a Verifier, verifying the quotes signature using the Intel SGX Provisioning Certification Service (PCS) and further inspecting the quote's included measurements \cite{IntelTDXWhitepaper2022}.%

\myparagraph{NVIDIA Hopper Architecture GPU} %
With NVIDIA's cGPU attestation, a relying party may verify an Attester GPU is indeed an NVIDIA-manufactured Hopper GPU running in CC-enabled mode. %
A CVM-contained relying party initiates the attestation by challenging the Attester cGPU with a nonce. %
This cGPU then generates an attestation report including evidence and the nonce, and signs this report with a private key termed the Attestation Key~(AK). %
This cGPU generates this attestation key along with a corresponding public key during its boot, deriving the key from boot device hardware measurements and finally signing the AK public key with a device-embedded device identity key.%

The cGPU returns the public keys of the AK and the device identity key to the Verifier, which form a certificate chain together with NVIDIA GPU Driver obtained certificates. %
The Verifier then checks this certificate chain while ensuring no certificates were revoked using the NVIDIA Online Certificate Status Protocol (OCSP)~\cite{NvidiaOCSPService} service.%

Thereupon, the Verifier receives and verifies the signature of the AK signed attestation report using the previously verified AK public key. %
Within this report, the Verifier obtains corresponding GPU driver and VBIOS identification, and uses this identification to request the corresponding Reference Integrity Manifest (RIM) files from the NVIDIA RIM service~\cite{NvidiaRIMService}. %
Further, the Verifier validates the RIM files and compiles a RIM files-derived  gold measurement list. %
The Verifier finally verifies the attestation report's measurements by comparing each report measurement to the Verifier-compiled golden measurements. %
If all report measurements match with a golden measurement, the GPU is assumed to be in an expected state \cite{gu_nvidia_2025, nertney2023confidential}.%

\subsection{Detailed Design}
\sys{} constitutes a secure confidential AI/ML system leveraging CPU and GPU Trusted Execution Environments (TEEs), protecting process-based applications interfacing with the GPU from privileged administrators with CVM access. %
To this end, \sys{} only provisions process-based applications with secrets after successful attestation and disables memory dumping in the Guest kernel. %
Hence, this architecture ensures the protection of sensitive data and computations throughout the ML lifecycle, including inference, training, and fine-tuning.%

To facilitate robust remote attestation of ML applications, a custom kernel module is implemented within the Guest OS of each confidential VM. This module performs dynamic measurements of running ML workloads, generating verifiable quotes that attest to the integrity and authenticity of the applications.%

\begin{itemize}
\item{\bf Configuration and Attestation Service (CAS):} Service acting as the root of trust for the entire system. CAS operates within its own TEE and can be directly attested by end users, ensuring transparency and verifiability. It manages security policies, handles attestation workflows, and provisions secrets (e.g., decryption keys) upon successful verification.
\item{\bf Kernel Module for Attestation:} Deployed in the Guest OS of confidential VMs, this module measures ML applications at runtime. It interacts with CAS to obtain signing keys, generates cryptographic quotes based on application hashes, and supports the overall remote attestation chain. In addition, we disables all memory-dumping capabilities within the Guest OS. The Guest OS itself is attested using the underlying TEE's attestation mechanism.
\item {\bf Security Policies:} User-defined configurations for the CAS that specify allowable measurements, attestation requirements, and secret provisioning rules for confidential VMs and ML applications.
\end{itemize}

\subsection{System Workflow}

\begin{figure}[t]
    \centering
    \includegraphics[scale=0.45]{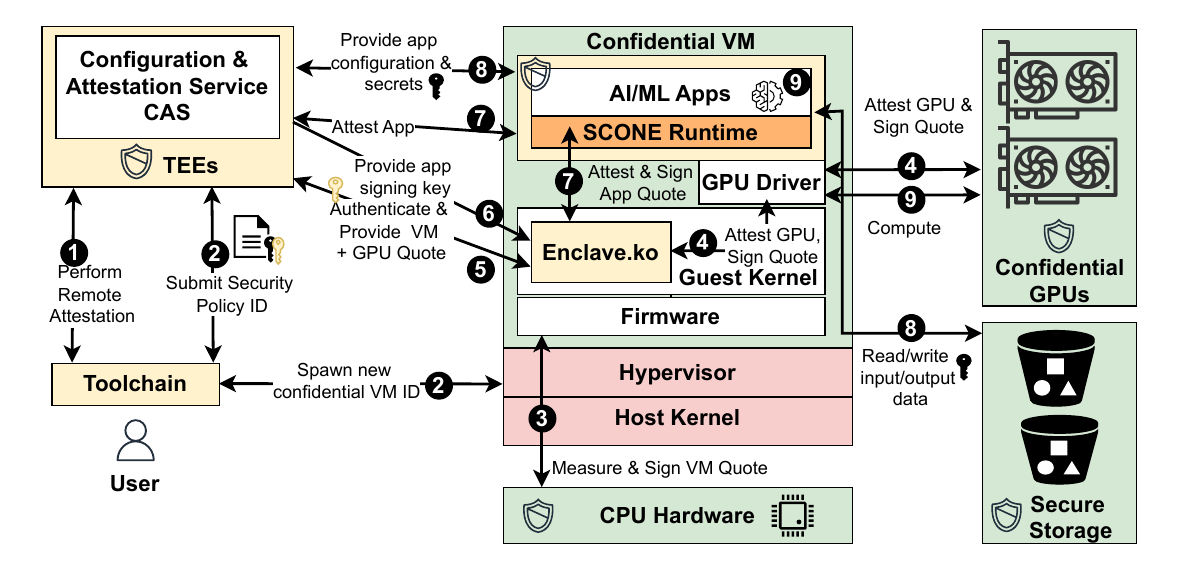}
    \caption{Secure AI/ML system design of \sys{} using SCONE}
    \label{fig:system_design}
\end{figure}

Figure~\ref{fig:system_design} illustrates the high-level workflow of the proposed system architecture.
The confidential AI/ML system workflow begins with the user defining a security policy for their AI/ML application and attesting the Configuration and Attestation Service (CAS)~\cite{palaemon} (step \blackcircled{1}), which operates within a Trusted Execution Environment (TEE) to establish trust. The user then submits the policy to CAS (step \blackcircled{2}). When spawning a confidential VM using Intel TDX, a VM-specific policy extension containing the VM ID is generated and uploaded to CAS (step \blackcircled{2}) following a re-attestation of CAS (step \blackcircled{3}). As the VM boots, its hardware measures and attests the firmware, which in turn attests the Guest OS and kernel, including a custom attestation kernel module. This module initiates a request to the GPU driver to attest confidential GPUs~\cite{nvidia_tee} (step \blackcircled{4}). The system then authenticates with CAS (step \blackcircled{5}), submitting attestation reports from both the confidential VM and GPUs. CAS uses these reports to issue a signing key according to the policy and marks the VM and GPUs as attested (step \blackcircled{6}). This guarantees the singleton property of the confidential VM~\cite{sinclave}. Next, the kernel module measures the running ML application—during inference or training—generates a cryptographic hash, signs it to produce a quote, and sends it to CAS (step \blackcircled{7}). CAS verifies the quote against the policy and, if valid, provisions configuration data and secrets such as decryption keys (step \blackcircled{8}), enabling the ML application to securely process data within the CPU and GPU TEE (step \blackcircled{9}).

%% file: 4-implementation.tex
\section{Implementation}

As we show in Figure \ref{fig:system_design}, we implement \sys using the SCONE framework~\cite{scone, palaemon} to run native AI/ML applications inside software-based enclaves without requiring any modifications to the application code. Regarding TEEs, we make use of Intel TDX for CVMs and NVIDIA H200 GPU for cGPUs. Our implementation relies on both Trustee~\cite{trustee} and SCONE attestation~\cite{palaemon} to build the \sys CAS: Trustee attests the CVM, while SCONE Attestation verifies the integrity of the AI/ML application itself.
To enable this functionality across diverse hardware platforms, we also developed the kernel module for attestation, referred to as the SCONE kernel module, which will be open sourced. This module allows \sys to operate on additional confidential computing architectures such as AMD SEV-SNP and ARM CCA, again without requiring changes to the AI/ML application code.

%% file: 5-evaluation.tex
\section{Benchmarks}

\begin{figure*}[t!]
    \centering
    \begin{subfigure}[t]{0.32\textwidth}
        \centering
        \includegraphics[width=\linewidth]{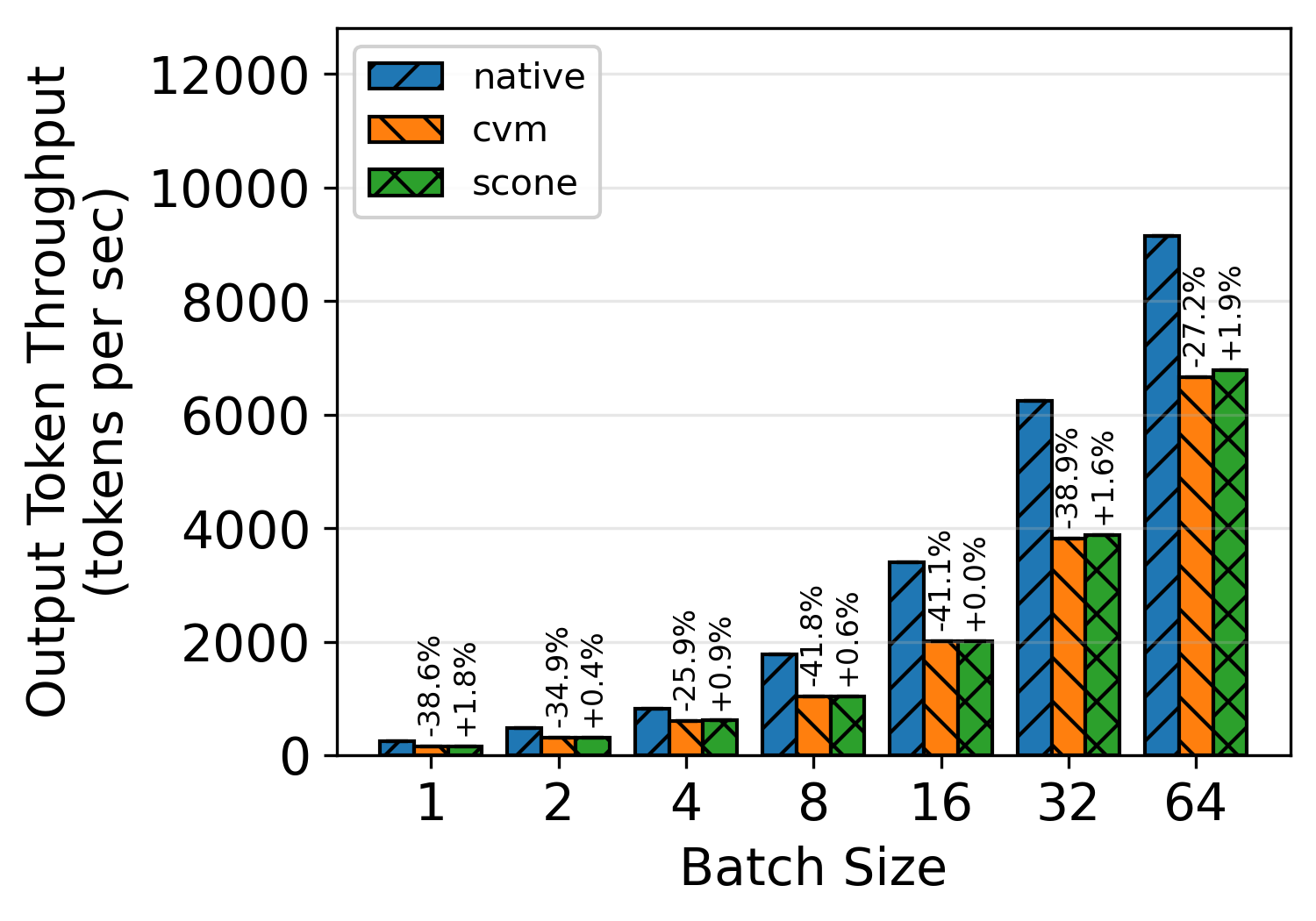}
        \caption{Throughput benchmark measured in Tokens Per Second (TPS). Batch size scaled from 1 to 64, with input- and output token size fixed to 128. SCONE to CVM overhead is negligible. CVM to Native overhead remains in the range of $62.8\%$ to $35.0\%$, decreasing with increased batch size.}
        \label{fig:tps_scaling}
    \end{subfigure}
    \hfill
    \begin{subfigure}[t]{0.32\textwidth}
        \centering
        \includegraphics[width=\linewidth]{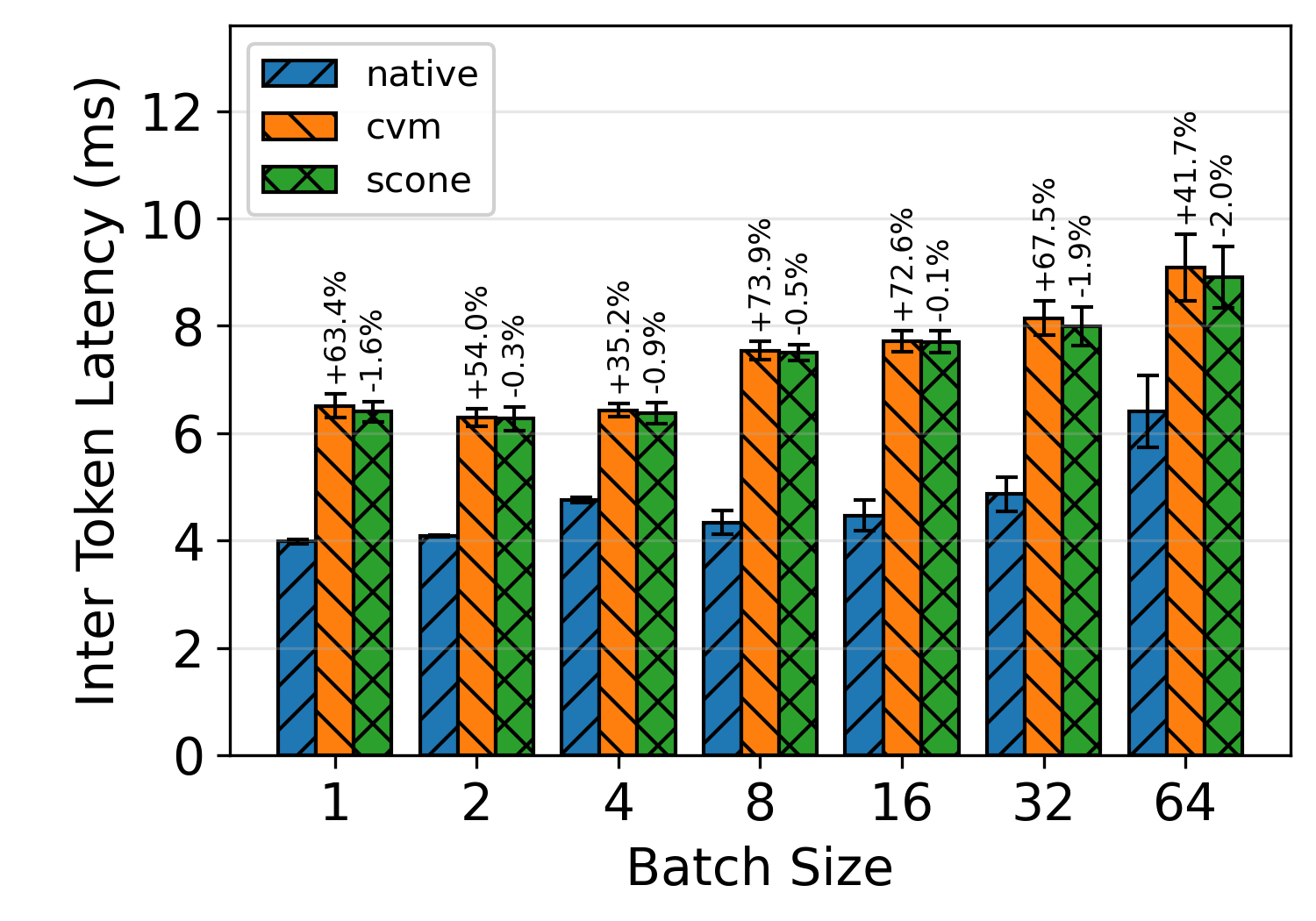}
        \caption{Inter-token Latency, i.e., Time Between Tokens (TBT) benchmark. Batch size scaled, with input- and output token size fixed to 128. SCONE to CVM overhead is negligible. CVM to Native overhead remains in the range of $73.9\%$ to $35.2\%$, decreasing with increased batch size.}
        \label{fig:tbt_scaling}
    \end{subfigure}
    \hfill
    \begin{subfigure}[t]{0.32\textwidth}
        \centering
        \includegraphics[width=\linewidth]{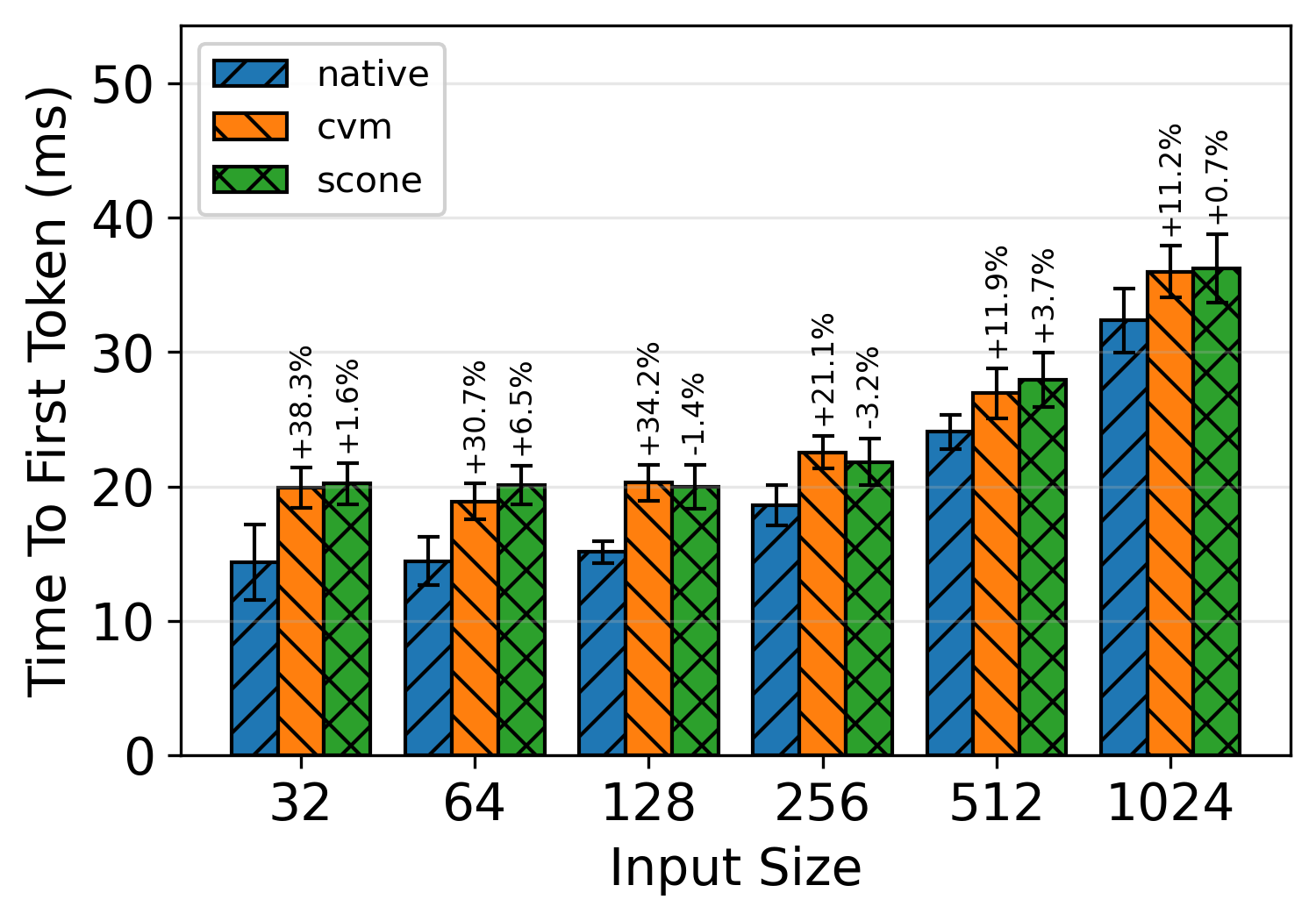}
        \caption{Time to First Token (TTFT) latency benchmark. Batch size is fixed to 1, output token size is 128, and input token size is scaled from 32 to 1024. SCONE to CVM overhead is negligible. CVM to Native overhead ranges from $38.3\%$ to $11.2\%$, decreasing with increased input token size.}
        \label{fig:ttft_scaling}
    \end{subfigure}

    \caption{LLM inference benchmarking results on the SYS-322GA-NR. We measure performance across three modes: (1) native, wherein the Guest is a native VM and the GPU is not in CC mode, (2) cvm, wherein the Guest is a TDX, the GPU is in CC mode, and (3) scone, consisting of the the \sys{}-modified TDX and GPU in CC mode. We issue $500\times{}batch\_size$ requests per benchmark, e.g., 32000 requests for batch size 64. The bar graphs show the absolute percentage difference relative to the predecessor bar (left to right).}
    \label{fig:main_experimental_results}
\end{figure*}
We evaluate \sys by benchmarking end-to-end workflows to confirm negligible overhead in LLM inference on H200 GPUs. Security validations include full attestation chain verification, guaranteeing production-grade reliability.%

\subsection{LLM Inference Performance}%
To determine \sys{}'s performance, we benchmark an \sys{}-enclosed LLM's inference. %
As a baseline, we run the LLM inference inside of a native Intel TDX CVM connected to the cGPU in confidential mode. %
We thereby measure Inter-Token Latency~(ITL) and the system throughput metric Tokens per Second~(TPS) while varying batch size. %
Further, we measure the Time-to-First-Token~(TTFT), thereby varying the input token size. %
For every benchmark, we issue $500\times{}batch\_size$ requests, e.g., for a batch size of 64, 32000 requests.%

Besides \sys{}'s overhead, we include native inference results of a non-TDX Guest with the H200 in non-CC mode. %
To this end, we display the confidential compute overhead of LLM inference.%

\myparagraph{Experimental Setup} %
We benchmark an environment-\hspace{0pt}contained NVIDIA-optimized \texttt{llama-3.1-8b-instruct}~\cite{llama3herd2024, nvidia_llama31_fp8_2024} LLM, serving the model with NVIDIA Triton Inference Server v25.01 with Triton v2.54.0~\cite{triton_server_2501} while using the TensorRT-LLM backend v0.17.0~\cite{NVIDIA_TensorRT_LLM_2023}. %
We further issue requests and measure the server performance using the NVIDIA \texttt{genai-\hspace{0pt}perf}~\cite{NVIDIA_GenAI_Perf_2024} benchmarking tool. %
We thereby optimize the model engine for the H200 by using FP8 for model weights and key-value cache. %
For the model parameters, we enable dynamic batching and set the key-value free GPU memory fraction to 0.95.%

Regarding environment, the hardware foundation is Supermicro's SYS-322GA-NR~\cite{sys322}, with dual Intel Xeon 6900-series processors (with P-cores up to 128 cores/256 threads per CPU and 500W TDP) and 755G of memory. %
The server is further equipped with an NVIDIA H200 NVL (141GB HBM3e) GPU, connected via PCIe 5.0 \cite{sys322}.%

We run the experiments on an Ubuntu 25.04 host, inside an Ubuntu 24.04.1 LTS Guest, hosted by QEMU 9.2.1 as the hypervisor, and starting the Guest with \texttt{100G} RAM and 32 vCPUs. %
We host the server with Docker v29.1.5 using the triton inference server image nvcr.io/nvidia/tritonserver:25.01-trtllm-python-py3.%

\myparagraph{RQ1} %
\textit{
What is \sys{}'s LLM-inference performance (TPS, TBT, and TTFT) overhead to native CVM's LLM inference performance?
} %
To measure TPS and TBT, we set input- and output token size to 128, while scaling the batch size from 1 up-to 64. %
To further measure TTFT, we set output token size to 128 and the batch size to 1, while scaling the input token size from 32 up-to 1024.%

\myparagraph{Results} %
As we show in Figures \ref{fig:tps_scaling}, \ref{fig:tbt_scaling}, and \ref{fig:ttft_scaling}, we do not measure any overhead for \sys{} compared to a native CVM deployment. %
Any small differences are insignificant and fall within the expected standard deviation.%

We achieve this performance by executing the SCONE-runtime in ``SIM'' mode; the runtime relies on the CVM memory protection and does not further encrypt memory. %
As such, the runtime executes syscalls synchronously without copying syscall arguments and delegates memory management to the kernel.%

\vspace{2mm}
\noindent\fbox{\parbox{\columnwidth}{
{\bf RQ1 takeaway:}
\sys{} imposes no performance overhead for LLM inference to the CVM baseline by only facilitating attestation and otherwise not interfering with the inference.%
}}

\myparagraph{RQ2} %
\textit{What is the native CC performance overhead compared to non-CC native inference?} %

\myparagraph{Results} %
Regarding TPS and TBT, we see a clear overhead of cvm to native in Figures \ref{fig:tps_scaling} and \ref{fig:tbt_scaling}. %
These vary between 35.0\% to 62.8\% overhead in TPS and between 35.2\% to 73.9\% overhead in TBT. %
In both benchmarks, the overheads tend to decrease with increased batch size. %
For TTFT, we likewise see an overhead of cvm to native in Figure \ref{fig:ttft_scaling}. %
This overhead ranges from $11.2\%$ to $38.3\%$, decreasing with increased input token size.%

We thus reproduce similar results as for the NVIDIA H100 in \cite{chrapek_confidential_2025} with the NVIDIA H200; with increasing batch- and input token size, the confidential compute cGPU performance penalty decreases. %
In particular, the cGPU confidential compute mode penalizes IO between CVM and cGPU, as this mode encrypts cGPU IO over bounce buffers between the CVM and the cGPU. %
Thus, every IO requires extra copies between trusted and untrusted memory. %
With increased batch- and input token size, the time of compute inside the cGPU increases, thus decreasing the impact of the confidential compute IO overhead. %
To this end, solutions such as Intel TDX-Connect~\cite{intel_tdx_connect} and AMD SEV-IO~\cite{amd_sev_tio} exist, yet NVIDIA has yet to support these technologies for its cGPUs~\cite{nvidia_trusted_solutions_notes}.%

\vspace{2mm}
\noindent\fbox{\parbox{\columnwidth}{
{\bf RQ2 takeaway:}
We show the same cGPU IO bottleneck to exist in the NVIDIA H200 as in the NVIDIA H100; cGPU IO overhead is high with small input token- and batch size, yet decreases with larger input token and batch sizes as the GPU compute time's impact increases.%
}}

\subsection{Attestation}%
We further investigate the attestation overhead of \sys{}. %
As a baseline, we select the native attestation latency of a TDX CVM including the H200 cGPU and further compare it to \sys{}'s attestation latency. %
Specifically, we measure the duration from the time the Attester initiates the attestation until the Attester receives the attestation result from the Verifier. %
This process consists of the Attester collecting evidence, the Attester sending the evidence to a Verifier, the Verifier comparing the evidence against an appraisal policy, and the Verifier returning the attestation result to the Attester.%

\myparagraph{Experimental Setup} %
We follow the setup of trustee~\cite{trustee} to facilitate remote attestation. %
With the CVM as the Attester and trustee as the Verfier, both share the same host to avoid network latency. %
Again, we use the SYS-322GA-NR as the hardware testbed, with TDX for the CVM and the NVIDIA H200 as the cGPU.%

Additional TD Quote generation and verification components, namely the Intel Provisioning Certificate Caching Service~(PCCS) and Quote Generation Service~(QGS), are also executed by the host OS to avoid additional latency. %
Furthermore, PCCS contains pre-cached collaterals from the Intel Provisioning Certificate Service~(PCS).%

Finally, we require minimal adjustments to trustee to enable the attestation of our system. %
Since our system is pre-production, the Intel Quote Verification Library~(QvL)~\cite{Intel_DCAP_2025} cannot verify our system's quote; the embedded Intel Root Public Key differs for pre-production quotes. 
As such, we skip over the QvL function for explicitly verifying the quote. %
Moreover, the trustee does not entirely support the verification of the H200 report measurements. %
We thus skip over unknown report measurements and verify all trustee-known measurements.%

To assess application‑level remote attestation, we enable the SCONE kernel module inside the Guest OS. This module requires a private key to sign attestation quotes. In a typical deployment, the key is obtained from the KBS/Trustee service after the CVM has been attested. For benchmarking, however, we provision this key manually using a dedicated provisioning process. We then create a minimal application, essentially a program that only executes {\em return 0}, so that the attestation step occurs before any real application logic runs. Along with this, we define the corresponding security policy and upload it to the SCONE CAS~\cite{palaemon}.

\myparagraph{RQ3} %
\textit{What overhead does \sys{} attestation add to native CVM and cGPU attestation?} %
We measure the latency of 100 attestations for both native and \sys{}, using the hyperfine~\cite{Peter_hyperfine_2025} tool as the benchmarking framework. %
For native, we use the trustee kbs-client tool to initiate the attestation within the CVM, including generating and collecting the TD Quote and cGPU SPDM measurements. %
For application‑level remote attestation, the minimal application is dynamically linked with the SCONE runtime so it can execute inside a software‑based enclave. The SCONE runtime measures the application to generate an attestation quote, which is then signed by the SCONE kernel module. Afterward, the signed quote is sent to the SCONE CAS for verification according to the defined security policy. We repeat this process 100 times to benchmark and record the attestation latency.

\myparagraph{Results} %
For the native TDX and cGPU attestation, we measure a mean latency of $1.1611$s with a standard deviation of $0.0313$s. %
We conducted additional benchmarking to assess application‑level remote attestation using the SCONE kernel module and SCONE runtime. The measurements yielded a mean latency of $0.0147$s with a standard deviation of $0.1508$s. %
Thus, the total \sys{} attestation takes around $1.1758$s, adding a negligible overhead of around $1.27$\%. %
SCONE CAS uses Edwards‑curve Digital Signature Algorithm (EdDSA) cryptography, e.g., Ed25519, for attestation, which enables very fast verification of attestation quotes.%

\vspace{2mm}
\noindent\fbox{\parbox{\columnwidth}{
{\bf RQ3 takeaway:}
\sys{} attestation adds an insignificant (\textasciitilde{}$1\%$) overhead when added onto CVM- and cGPU attestation.%
}}

%% file: 6-conclusion.tex
\subsection{Conclusion}
\sys{} minimizes the trusted computing base (TCB) by relying on hardware-enforced TEEs and granular attestation, reducing the attack surface while supporting unmodified ML applications. %
Specifically, \sys{} enables CVM-enclosed process-based TEEs which protect against K8s-admins with access to the CVM. %
The system further enables secure multi-tenant cloud deployments for sensitive workloads in domains like healthcare, finance, and research. Future enhancements could include support for additional TEE technologies or federated learning scenarios to further extend confidentiality across distributed environments.%